\documentclass[12pt]{iopart}
\usepackage{graphicx}
\usepackage{amsfonts}
\usepackage{amssymb}
\begin{document}
\title{Aspects of finite electrodynamics in $D=3$ dimensions}
\author{Patricio Gaete\dag\footnote{e-mail address: patricio.gaete@usm.cl},
Jos\'{e} Helay\"{e}l-Neto\ddag\footnote{e-mail address: helayel@cbpf.br}
Euro Spallucci\ddag\footnote{e-mail address: spallucci@ts.infn.it}}
\address{\dag\ Departamento de F\'{\i}sica and Centro Cient\'{i}fico-Tecnol\'ogico de Valpara\'{i}so, 
Universidad T\'ecnica Federico Santa Mar\'{\i}a, Valpara\'{\i}so, Chile}
\address{\ddag\ Centro Brasileiro de Pesquisas F\'{i}sicas (CBPF),
Rio de Janeiro, RJ, Brazil}
\address{\ddag\ Dipartimento di Fisica Teorica, Universit\`a di \
Trieste and INFN, Sezione di Trieste, Italy}

\begin{abstract}
We study the impact of a minimal length on physical observables for a
three-dimensional axionic electrodynamics. Our calculation is done
within the framework of the gauge-invariant, but path-dependent, variables formalism which is alternative to the Wilson loop approach. 
Our result shows that the interaction energy contains a regularized Bessel function
and a linear confining potential. This calculation involves no $\theta$ expansion at all. Once again, the present analysis displays the key role
played by the new quantum of length.\end{abstract}
\pacs{14.70.-e, 12.60.Cn, 13.40.Gp}
\submitted
\maketitle

\section{Introduction}

One of the most actively pursued areas of research in high energy physics consists of the investigation of extensions of the Standard Model (SM). 
This is primarily because the SM does not include a quantum theory of gravitational interactions. 
As is well known, in the search for a more fundamental theory going beyond the SM string theories are
the only known candidate for a consistent, ultraviolet finite quantum theory
of gravity, unifying all fundamental interactions. It should, however, be noted here that string theories apart from the metric also predict 
the existence of a scalar field (dilaton), an antisymmetric tensor field of the third rank which is associated with torsion and noncommutativity. 
This has led to an increasing interest in possible physical effects of noncommutativity in quantum field theories, which have been studied using 
the Moyal star-product \cite{Witten:1985cc,Seiberg:1999vs,Douglas:2001ba,Szabo:2001kg,Gomis:2000sp,Bichl:2001nf}. Mention should be made, at this 
point,  to a novel way to formulate noncommutative quantum field theory (or quantum field theory in the presence of a minimal length) 
\cite{Smailagic:2003rp,Smailagic:2003yb,Smailagic:2004yy} which
clearly leads to an ultraviolet finite field theory and the cutoff is provided
by the noncommutative parameter $\theta$. In this connection, it may be
recalled that the essential idea of this development is to define the fields as
mean value over coherent states of the noncommutative plane, such that a star product needs not be introduced. We further note that recently 
it has been shown that the coherent state approach can be summarized through the
introduction of a new multiplication rule which is known as Voros
star-product \cite{Galluccio:2008wk,Galluccio:2009ss},
\cite{Banerjee:2009xx,Gangopadhyay:2010zm,Basu:2011kh}. Nevertheless, and most importantly, physics turns out be independent from the choice of 
the type of product \cite{Hammou:2001cc}. 

On the other hand, it is well known that a full understanding of the QCD vacuum structure and color confinement mechanism from first principles 
remain still elusive. However, phenomenological models have been of importance in our present understanding of confinement, and can be 
considered as effective theories of QCD. It is worthy recalling here that many approaches to the problem of confinement rely on the phenomenon 
of condensation. For example, in the illustrative scenario of dual superconductivity 
\cite{Nambu:1974zg,'t Hooft,Mandelstam:1974pi} the condensation is due to topological defects originated from quantum fluctuations (monopoles). 
Accordingly, the color electric flux linking quarks is squeezed into strings (flux tubes), and the nonvanishing string tension represents 
the proportionality constant in the linear, quark confining, potential. In this respect, it is appropriate to recall that Abelian gauge theories 
also possess a confining phase, by including the effects due to the compactness of the U(1) group, which dramatically changes the infrared 
properties of the model \cite{Polyakov:1976fu}. These results, first found in \cite{Polyakov:1976fu}, have been ever since recovered by many 
different techniques 
\cite{Ezawa:1979qy,Orland:1981ku,Kondo:1998nw}  where the key ingredient is the contribution of self-dual topological excitations.

With these ideas in mind, in a previous paper \cite{Gaete:2011ka}, we have studied axionic electrodynamics from this new noncommutative 
approach (coherent state approach), in the presence of a nontrivial constant
expectation value for the gauge field strength. In particular, in the case of a constant magnetic field strength expectation value, 
we have obtained an ultraviolet finite static potential which is the sum of a Yukawa-type and a linear potential, leading to the confinement of 
static charges. We note that this theory experiences mass generation due to the breaking of rotational invariance induced by the 
classical background configuration of the gauge field strength. Interestingly, it should be noted that this calculation involves 
no $\theta$ expansion at all. By following this line of reasoning, the present work is aimed at studying the stability of the above scenario 
for the three-dimensional case. The main purpose here is to reexamine the effects of this new noncommutativity on a physical observable, and 
to check if a linearly increasing gauge potential is still present whenever we go over into three dimensions. 

At this point, we would like to recall that three-dimensional theories are interesting because of their
 connection to the high-temperature 
limit of four-dimensional theories \cite{Appelquist:1981vg,Jackiw:1980kv,Lee:1993mv, Das}, as well as, for their applications to condensed 
matter physics \cite{Stone}. Most recently, three-dimensional physics has been raising a great deal of interest in connection with branes study,
namely,issues like self-duality and new possibilities for supersymmetry breaking as induced by $3$-branes are of special relevance.

Thus, as already mentioned, the main purpose here is to examine the 
effects of this new noncommutativity on a physical observable for the three-dimensional case. To do this, we will work out the static potential 
for axionic
electrodynamics by using the gauge-invariant but path-dependent variables formalism along the lines of 
Refs. \cite{Gaete:2004ga, Gaete:2007ax, Gaete:2007zn}. According to this formalism the gauge fixing procedure corresponds to a path choice. 
Nevertheless, the point we wish to emphasize is that this approach offers a natural setting to examine aspects of screening and confinement in 
gauge theories, because it involves the use of strings to carry electric flux \cite{Gaete:1997eg}. As we will see, there are two generic 
features that are common in the four-dimensional case and its lower extension studied here. First, the existence of a linear potential, 
leading to the confinement of static charges. The second point is related to the correspondence among diverse effective theories. 
In fact, in the case of a constant magnetic field strength expectation value, we obtain that the interaction energy is the sum of a 
regularized Bessel function and a linear potential. Incidentally, the above static potential profile is analogous to that encountered in: 
a Lorentz-and CPT- violating Maxwell-Chern-Simons model \cite{Belich:2002vd}, a Maxwell-like three-dimensional model induced by the 
condensation of topological defects driven by quantum fluctuations \cite{Gaete:2005ht}, a Lorentz invariant violating electromagnetism arising 
from a Julia-Toulouse mechanism \cite{Gaete:2006ss}, and three-dimensional gluodynamics in curved space-time \cite{Gaete:2007sj}.

\section{Three-dimensional finite electrodynamics}
\subsection{Maxwell case}

As already mentioned, our principal purpose is to calculate explicitly the interaction energy between static point-like
sources for non-commutative axionic electrodynamics. However, before going into this theory, we shall discuss the interaction energy 
for non-commutative electrodynamics, through two different methods. The first approach is based on the path-integral formalism, 
whereas the second one makes use of the gauge-invariant but path-dependent variables formalism. This would not only provide the 
setup for our subsequent work, but also fix the notation. \\

The starting point is the three-dimensional, imaginary time rotated, euclidean Lagrangian: 
\begin{equation}
\mathcal{L} =
- \frac{1}{4}F_{\mu \nu }\, F^{\mu \nu }. \label{NCthreed5}
\end{equation}
The effects of a minimal length $l_0\equiv \sqrt\theta $   induced by ``fluctuating'' non-commutative coordinates 
 can be described in various ways.  The most common approach is to shift non-commutativity from coordinates to
the product of functions (fields) by introducing the so-called ``star-product''.  Unfortunately, once star-multiplication
is introduced the only way to carry out calculations is through perturbative expansion in $\theta$ leading to inconsistent
results (see the Appendix in \cite{Gaete:2011ka}) \footnote{A remarkable exception is provided by some cases 
where it is possible to map the formulation with the
star product onto matrix models, which can be simulated numerically 
\cite{Bietenholz:2002ch,Bietenholz:2004xs,Bietenholz:2006cz,Bietenholz:2007ia}}. 
An alternative, non-perturbative approach, avoiding expansion in $\theta$,
has been introduced in \cite{Smailagic:2003rp,Smailagic:2003yb,Smailagic:2004yy} and turned out to introduce a simple modification
in the Feynman propagators as final result.
Let us then write down the functional generator of the Green's functions, that is,
\begin{equation}
Z\left[ J \right] = \exp \left( { - \frac{1}{2}\int {d^3 xd^3 y}
J^\mu \left( x \right)D_{\mu \nu } \left( {x,y} \right)J^\nu  \left(
y \right)} \right). \label{NCthreed10}
\end{equation}
Next, adding to (\ref{NCthreed5}) the gauge-fixing term
${\cal{L}}_{GF} = -\frac{1}{2}(\partial_\mu A^\mu)^2$ (Feynman
gauge), and noting that no Faddeev-Popov ghosts are required in this
case, we  get the propagator in momentum space
\begin{equation}
D_{\mu \nu } \left( k \right) =   \frac{1}{{k^2 }}\left\{
{e^{-\theta k^2 } \delta _{\mu \nu }  + \left( {1 - e^{-\theta k^2 } }
\right)\frac{{k_\mu  k_\nu  }}{{k^2 }}} \right\}. \label{NCthreed15}
\end{equation}
Equation (\ref{NCthreed15}) shows as only short wavelength are  suppressed
by the underlying space(time) quantum fluctuations.  The exponential damping factor in the propagator can be
seen as a \emph{friction} effect experienced by the photon at length scale comparable with $\sqrt{\theta}$, where
the classical model of space(time) as a smooth manifold breaks down. Deviations from classical behavior can be
seen as an increasing degree of ``roughness'', or fuzzyness, opposing to photon propagation.\\
By means of expression $Z = e^{-W\left[ J \right]}$, and employing
Eq. (\ref {NCthreed15}), $W\left[ J \right]$ takes the form
\begin{equation}
W\left[ J \right] =  \frac{1}{2}\int {\frac{{d^3 k}}{{\left( {2\pi
} \right)^3 }}} J_\mu ^ *  \left( k \right)\left [ { 
\frac{{e^{-\theta k^2 } }}{{k^2 }}\delta ^{\mu \nu } - \frac{{\left( {1
- e^{-\theta k^2 } } \right)}}{{k^2 }}\frac{{k^\mu  k^\nu  }}{{k^2
}}} \right ]J_\nu  \left( k \right). \label{NCthreed20}
\end{equation}
A \emph{bonus} of our approach is that, even if $\theta $ has dimension of length squared, or mass to power minus two, 
the way it enters the propagator does not break gauge invariance and  the current $J^\mu (k)$ is still 
divergence-free. Thus,  expression (\ref{NCthreed20})  becomes
\begin{equation}
W\left[ J \right] = \frac{1}{2}\int {\frac{{d^3 k}}{{\left( {2\pi }
\right)^3 }}} J_\mu ^ *  \left( k \right)\left( {\frac{{e^{-\theta
k^2 } }}{{k^2 }}} \right)J^\mu  \left( k \right). \label{NCthreed25}
\end{equation}
Next, for $J_\mu  \left( x \right) = \left[ {Q\delta ^{\left( 2
\right)} \left( {{\bf x} - {\bf x}^{\left( 1 \right)} } \right) + Q^
\prime  \delta ^{\left( 2 \right)} \left( {{\bf x} - {\bf x}^{\left(
2 \right)} } \right)} \right]\delta _\mu ^0$, and using standard
functional techniques \cite{Zee}, we obtain that the interaction
energy of the system is given by
\begin{equation}
V\left( r \right) = QQ^ \prime  \int {\frac{{d^2 k}}{{\left( {2\pi }
\right)^2 }}\frac{{e^{ - \theta {\bf k}^2 } }}{{{\bf k}^2 }}}
e^{i{\bf k} \cdot {\bf r}} , \label{NCthreed30}
\end{equation}
where $ {\bf r} \equiv {\bf x}^{\left( 1 \right)}  - {\bf x}^{\left(
2 \right)}$. 

Now, we move on to calculate the integral (\ref{NCthreed30}). To this end it is advantageous to introduce an infrared regulator $\mu$. 
This allows us to obtain a form more comfortable to handle the integral. Hence we evaluate 
$\mathop {\lim }\limits_{\varepsilon  \to 0} \tilde I$, that is,
\begin{eqnarray}
I \equiv \mathop {\lim }\limits_{\varepsilon  \to 0} \tilde I &=& \mathop {\lim }\limits_{\varepsilon  \to 0} \left( {\mu ^2 } 
\right)^{ - {\raise0.7ex\hbox{$\varepsilon $} \!\mathord{\left/
 {\vphantom {\varepsilon  2}}\right.\kern-\nulldelimiterspace}
\!\lower0.7ex\hbox{$2$}}} \int {\frac{{d^{2 + \varepsilon } k}}{{\left( {2\pi } \right)^2 }}} 
\frac{{e^{ - \theta {\bf k}^2 } }}{{{\bf k}^2 }}e^{i{\bf k} \cdot {\bf r}} \nonumber\\ 
&=& \mathop {\lim }\limits_{\varepsilon  \to 0} \left( {\mu ^2 } \right)^{ - {\raise0.7ex\hbox{$\varepsilon $} \!\mathord{\left/
 {\vphantom {\varepsilon  2}}\right.\kern-\nulldelimiterspace}
\!\lower0.7ex\hbox{$2$}}} \int_0^\infty  {ds} \int {\frac{{d^{2 + \varepsilon } k}}{{\left( {2\pi } \right)^2 }}} 
e^{ - (\theta  + s){\bf k}^2 }e^{i{\bf k} \cdot {\bf r}} .
\label{NCthreed35a}
\end{eqnarray}
We may further simplify Eq.(\ref{NCthreed35a}) by doing the ${\bf k}$ and $s$ integral, which leads immediately to the result 
\begin{equation}
I = \frac{1}{{4\pi }}\mathop {\lim }\limits_{\varepsilon  \to 0} 
\left( {\mu ^2 r^2 } \right)^{ - {\raise0.7ex\hbox{$\varepsilon $} \!\mathord{\left/
 {\vphantom {\varepsilon  2}}\right.\kern-\nulldelimiterspace}
\!\lower0.7ex\hbox{$2$}}} \gamma \left( {{\raise0.5ex\hbox{$\scriptstyle \varepsilon $}
\kern-0.1em/\kern-0.15em
\lower0.25ex\hbox{$\scriptstyle 2$}},{\raise0.5ex\hbox{$\scriptstyle {r^2 }$}
\kern-0.1em/\kern-0.15em
\lower0.25ex\hbox{$\scriptstyle {4\theta }$}}} \right). \label{NCthreed35b}
\end{equation}
Here $\gamma \left(
{{\raise0.5ex\hbox{$\scriptstyle \varepsilon$}
\kern-0.1em/\kern-0.15em\lower0.25ex\hbox{$\scriptstyle 2$}};
{\raise0.5ex\hbox{$\scriptstyle {r^2 }$}\kern-0.1em/\kern-0.15em
\lower0.25ex\hbox{$\scriptstyle {4\theta }$}}} \right)$ is the lower
incomplete Gamma function defined by the following integral
representation
\begin{equation}
\gamma \left(\, \frac{a}{b}\ ; x\, \right) \equiv \int_0^x
{\frac{{du}}{u}}\, u^{{a \mathord{\left/ {\vphantom {a b}} \right.
\kern-\nulldelimiterspace} b}}\, e^{ - u}. \label{NCthreed40}
\end{equation}
Next, we use $\gamma \left( {{\raise0.5ex\hbox{$\scriptstyle \varepsilon $}
\kern-0.1em/\kern-0.15em
\lower0.25ex\hbox{$\scriptstyle 2$}},{\raise0.5ex\hbox{$\scriptstyle {r^2 }$}
\kern-0.1em/\kern-0.15em
\lower0.25ex\hbox{$\scriptstyle {4\theta }$}}} \right) = 
\frac{2}{\varepsilon }\left[ {\left( {\frac{{r^2 }}{{4\theta }}} \right)^{{\raise0.5ex\hbox{$\scriptstyle \varepsilon $}
\kern-0.1em/\kern-0.15em
\lower0.25ex\hbox{$\scriptstyle 2$}}} e^{ - {\raise0.7ex\hbox{${r^2 }$} \!\mathord{\left/
 {\vphantom {{r^2 } {4\theta }}}\right.\kern-\nulldelimiterspace}
\!\lower0.7ex\hbox{${4\theta }$}}}  + \gamma \left( {1 + {\raise0.5ex\hbox{$\scriptstyle \varepsilon $}
\kern-0.1em/\kern-0.15em
\lower0.25ex\hbox{$\scriptstyle 2$}},{\raise0.5ex\hbox{$\scriptstyle {r^2 }$}
\kern-0.1em/\kern-0.15em
\lower0.25ex\hbox{$\scriptstyle {4\theta }$}}} \right)} \right]
$, $\left( {\mu ^2 r^2 } \right)^{ - {\raise0.5ex\hbox{$\scriptstyle \varepsilon $}
\kern-0.1em/\kern-0.15em
\lower0.25ex\hbox{$\scriptstyle 2$}}}  \to 1 - \frac{\varepsilon }{2}\ln \left( {\mu ^2 r^2 } \right)
$,  $\left(\, \frac{\textstyle{r^2 }}{4\theta}\, \right)^{{\raise0.5ex\hbox{$\scriptstyle \varepsilon $}
\kern-0.1em/\kern-0.15em
\lower0.25ex\hbox{$\scriptstyle 2$}}}  \to 1 + \frac{\textstyle{\varepsilon}}{2}\ln \left( \frac{\textstyle{r^2 }}{4\theta} \right)$ and $  
\gamma \left( {1 + {\raise0.5ex\hbox{$\scriptstyle \varepsilon $}
\kern-0.1em/\kern-0.15em
\lower0.25ex\hbox{$\scriptstyle 2$}},{\raise0.5ex\hbox{$\scriptstyle {r^2 }$}
\kern-0.1em/\kern-0.15em
\lower0.25ex\hbox{$\scriptstyle {4\theta }$}}} \right) \to \gamma \left( {1,{\raise0.5ex\hbox{$\scriptstyle {r^2 }$}
\kern-0.1em/\kern-0.15em
\lower0.25ex\hbox{$\scriptstyle {4\theta }$}}} \right) = 1 - e^{ - {\textstyle{{r^2 } \over {4\theta }}}} $, to examine the behavior 
of expression
(\ref{NCthreed35b}) as $\varepsilon  \to 0$. Expression (\ref{NCthreed35b}) 
then becomes
\begin{equation}
I=- \frac{1}{2\pi}\, 
\left[\,\ln\left(\,\mu r\,\right)  - e^{-r^2/4\theta} \ln\left(\,\frac{ r}{2\sqrt\theta}\,\right) \,\right]
\label{NCthreed45}
\end{equation}
Combining Eqs. (\ref {NCthreed30}) and (\ref {NCthreed45}), together with $-Q=Q^\prime$, the interaction energy reduces to 
\begin{equation}
V\left(\, r\, \right) =  
\frac{Q^2}{2\pi}\, 
\left[\,\ln\left(\,\mu r\,\right)  - e^{-r^2/4\theta} \ln\left(\,\frac{ r}{2\sqrt\theta}\,\right) \,\right]
\label{NCthreed55}
\end{equation}
It is interesting to notice that
unlike the Coulomb potential which is singular at the origin,
$V$ is finite there: $V\left( 0 \right) = {{Q^2 } \mathord{\left/
 {\vphantom {{Q^2 } {2\pi }}} \right.
 \kern-\nulldelimiterspace} {2\pi }}\ln \left( {2\mu \sqrt \theta  } \right)
$. The fact that the potential is finite for $r \rightarrow 0$, it is a clear evidence
that the self-energy and the electromagnetic mass of a point-like
particle are finite in this noncommutative version of electrodynamics. 
However, when $r$ is large, $V$ reduces to the Coulomb potential (Fig.1).

\begin{figure}[h]
\begin{center}
\includegraphics[scale=1.42]{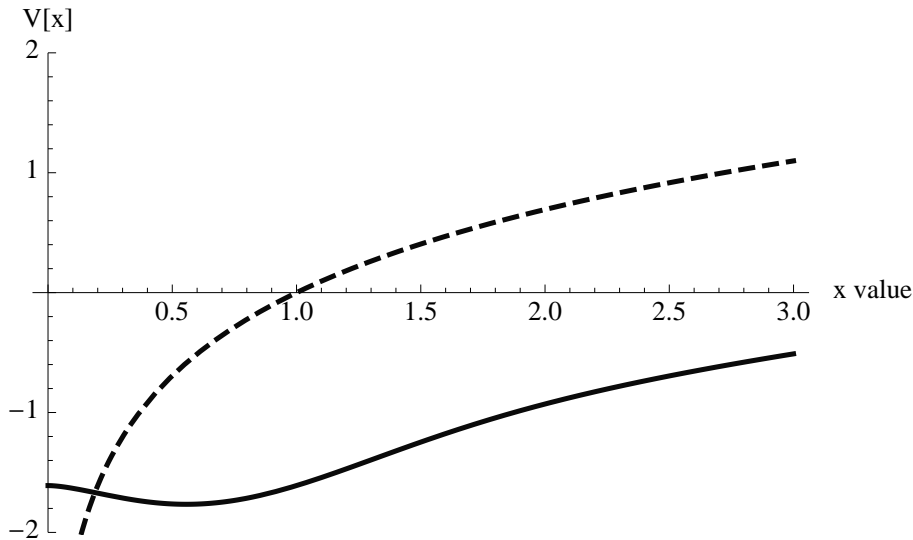}
\end{center}
\caption{\small The potential $V$ (in units of $\frac{Q^2}{2 \pi}$),
as a function of the distance $r$. The dashed line represents the
Coulomb potential (in units of $\frac{Q^2}{2 \pi}$).\label{fig1}}
\end{figure}

Next we compute the interaction energy from the viewpoint of the
gauge-invariant but path-dependent variables formalism, along the
lines of Refs. \cite{Gaete:2004ga,Gaete:2007ax,Gaete:2007zn}. Within
this framework, we shall compute the expectation value of the energy
operator $H$ in the physical state $|\Phi\rangle$, which we will
denote by ${\langle H \rangle}_\Phi$.

By proceeding in the same way as in \cite{Gaete:2011ka}, we obtain the static potential for two opposite charges, 
located at ${\bf y}$ and ${\bf y^{\prime}}$:

\begin{equation}
V(L) = \frac{{Q^2 }}{{2\pi }}
\left[\,\ln\left(\,\mu L\,\right)  - e^{-L^2/4\theta} \ln\left(\,\frac{ L}{2\sqrt\theta}\,\right) \,\right]
\label{NCthreed110}
\end{equation}
with $|{\bf y} -{\bf y}^{\prime}|\equiv L$. It  is remarkable that
two quite different methods have led to the same expression for the
effective three-dimensional potential. This astonishing result seems
to indicate that to lower orders the two approaches  might be
equivalent order by order. 

\subsection{Maxwell-Chern-Simons case}

We now consider the calculation of the interaction energy between static
point-like sources in a topologically massive gauge theory. In such a case
the Lagrangian reads \cite{Deser:1981wh}: 
\begin{equation}
{\cal L} = -\frac{1}{4}F_{\mu \nu }^{2}+\frac{\sigma}{4}
\varepsilon ^{\mu \nu \rho }A_{\mu }F_{\nu \rho }-A_{0}J^{0},  \label{NCthreed140}
\end{equation}
where $J^{0}$ is the external current and $\sigma$ is the topological mass.

The above Lagrangian will be the starting point of the Dirac constrained analysis. The canonical momenta following from 
Eq. (\ref{NCthreed140}) are
$\pi ^{\mu}=-F^{0\mu }+\frac{\sigma }{2}\varepsilon ^{0\mu \nu }A_{\nu }$,
which results in the usual primary constraint $\pi ^{0}=0$ and $\pi^{i}=F^{i0}+\frac{\sigma}{2}\varepsilon ^{ij}A_{j}$ $(i,j=1,2).$ 
So the canonical Hamiltonian is 
\begin{equation}
H_{c}=\int d^{2}x\left(\pi_{i}\partial ^{i}A_{0} -\frac{1}{2}F_{i0}F^{i0}+\frac{1}{4}F^{ij}F_{ij} - \frac{\sigma}{2}
\varepsilon ^{ij}A_{0}\partial
_{i}A_{j}+A_{0}J^{0}\right) .  \label{NCthreed145}
\end{equation}
Time conservation of the constraint $\pi ^{0}$ leads to the secondary
constraint (Gauss law) $\Omega _{1}\left( x\right) =\partial _{i}\pi ^{i}+\frac{\sigma}{2} \varepsilon _{ij}\partial ^{i}A^{j}-J^{0}=0$, 
and the time stability of the secondary constraint does not induce more constraints, 
which are first class. It should be noted that the constrained structure
for the gauge field remains identical to the Maxwell theory. Thus, the
quantization can be done in a similar manner to that in the previous
subsection. In view of this situation, and in order to illustrate the discussion, we now write the equations of motion in terms of 
the magnetic $\left( B=\varepsilon _{ij}\partial ^{i}A^{j}\right) $ and electric $(E^{i}=\pi ^{i}-\frac{\sigma}{2} \varepsilon ^{ij}A_{j})$ 
fields as 
\begin{equation}
\dot E_i \left( x \right) =  - 2\sigma \varepsilon _{ij} E^j \left( x \right) - \varepsilon _{ij} \partial ^j B,  \label{NCthreed150a}
\end{equation}
\begin{equation}
\dot B\left( x \right) =  - \varepsilon _{ij} \partial ^i E^j.  \label{NCthreed150b}
\end{equation}
In the same way, we write the Gauss law as 
\begin{equation}
\partial _{i}E_{L}^{i}+\sigma B-J^{0}=0,  \label{NCthreed155}
\end{equation}
where $E_{L}^{i}$ refers to the longitudinal part of $E^{i}$.
This implies that for a static charge located at $x^{i}=0$, the static
electromagnetic fields are given by
\begin{equation}
B =  - \sigma \frac{{J^0 }}{{\nabla ^2  - \sigma ^2 }},  \label{NCthree160}
\end{equation}
\begin{equation}
E_i \left( x \right) = \frac{1}{\sigma }\partial _i B, \label{NCthreed165}
\end{equation}
where $\mathbf{\nabla }^{2}$ is the two-dimensional Laplacian. 
For $ J^0 \left( x \right) = qe^{\theta \nabla ^2 } \delta ^{\left( 2 \right)} \left( {{\bf x}} \right)$, 
expressions (\ref{NCthree160}) and (\ref{NCthreed165}) reduce to
\begin{equation}
B = q\sigma \frac{{e^{\sigma ^2 \theta } }}{{2\pi }}\left\{ {K_0 \left( {\sigma r} \right)
 - \frac{1}{2}\int\limits_{{\raise0.5ex\hbox{$\scriptstyle r$}
\kern-0.1em/\kern-0.15em
\lower0.25ex\hbox{$\scriptstyle {2\sigma \theta }$}}}^\infty  {dy} \frac{1}{y}e^{ - {\textstyle{{\sigma r} \over 2}}
\left( {y + \frac{1}{y}} \right)} } \right\},  \label{NCthreed170a}
\end{equation}
\begin{equation}
E_i  = q\frac{{e^{\sigma ^2 \theta } }}{{2\pi }}\partial _i \left\{ {K_0 \left( {\sigma r} \right)
 - \frac{1}{2}\int\limits_{{\raise0.5ex\hbox{$\scriptstyle r$}
\kern-0.1em/\kern-0.15em
\lower0.25ex\hbox{$\scriptstyle {2\sigma \theta }$}}}^\infty  {dy} \frac{1}{y}e^{ - {\textstyle{{\sigma r} \over 2}}
\left( {y + \frac{1}{y}} \right)} } \right\},  \label{NCthreed170b}
\end{equation}
where $r = |{\bf x}|$ and $K_{0}$ is the modified Bessel's function.

Having made these observations, we can write immediately the following expression for the physical scalar potential \cite{Gaete:2011ka}: 
\begin{equation}
{\cal A}_0 \left( {t,x} \right) = \int_0^1 {d\lambda } x^i E_i \left( {t,\lambda {\bf x}} \right) = \int_0^1 {d\lambda } x^i 
\partial _i^{\lambda {\bf x}} \left( { - \frac{{J^0 \left( {\lambda {\bf x}} \right)}}{{\nabla ^2  - \sigma ^2 }}} \right), \label{NCthreed175}
\end{equation}
For $J^o \left( {\bf x} \right) = qe^{\theta \nabla ^2 } \delta ^{\left( 2 \right)} \left( {{\bf x} - {\bf a}} \right)$ expression 
(\ref{NCthreed175}) then becomes 
\begin{eqnarray}
 {\cal A}_0 \left( {\bf x} \right) &=& q\frac{{e^{\sigma ^2 \theta } }}{{2\pi }}\left\{ {K_0 \left( {\sigma |{\bf x} - {\bf a}  } \right)| 
- \frac{1}{2}\int\limits_{{\raise0.5ex\hbox{$\scriptstyle {|{\bf x} - {\bf a}|}$}
\kern-0.1em/\kern-0.15em
\lower0.25ex\hbox{$\scriptstyle {2\sigma \theta }$}}}^\infty  {dt\frac{1}{t}e^{ - |{\bf x} - {\bf a}|
\left( {t + {\textstyle{1 \over t}}} \right)} } } \right\} \nonumber\\ 
&-& q\frac{{e^{\sigma ^2 \theta } }}{{2\pi }}\left\{ {K_0 \left( {\sigma
{|{\bf a}|}} \right) - \frac{1}{2}\int\limits_{{\raise0.5ex\hbox{$\scriptstyle {|{\bf a}|}$} \kern-0.1em/\kern-0.15em 
\lower0.25ex\hbox{$\scriptstyle {2\sigma \theta }$}}}^\infty  {dt\frac{1}{t}e^{ - |{\bf a}|\left( {t + {\textstyle{1 \over t}}} \right)} } } 
\right\}. \label{NCthreed180}
\end{eqnarray}

As we have explained in \cite{Gaete:2011ka}, the interaction energy for a pair of static point-like opposite charges at $\bf y$ 
and $\bf {y}^{\prime }$ is given by 
\begin{equation}
V(|{\bf y} - {\bf y}^ {\prime}|)  = - q^2 \frac{{e^{\sigma ^2 \theta } }}{2\pi }\left\{ {K_0 \left( {\sigma | {\bf y} - {\bf y}^ {\prime}  |} 
\right) - \frac{1}{2}\int\limits_{{\raise0.5ex\hbox{$\scriptstyle {| {\bf y} - {\bf y}^ {\prime}  |}$}
\kern-0.1em/\kern-0.15em
\lower0.25ex\hbox{$\scriptstyle {2\sigma \theta }$}}}^\infty  {dt\frac{1}{t}e^{ - {\textstyle{\sigma  \over 2}}|{\bf y} - {\bf y}^ {\prime}  |
\left( {t + \frac{1}{t}} \right)} } } \right\},
\label{NCthreed185}
\end{equation}
which is ultraviolet finite (Fig.2). Note that in Fig. (2) we defined 
$V(|{\bf y} - {\bf y}^ {\prime}|) = q^2 \frac{{e^{\sigma ^2 \theta } }}{{2\pi }}V\left[ x \right]$.

\begin{figure}[h]
\begin{center}
\includegraphics[scale=1.42]{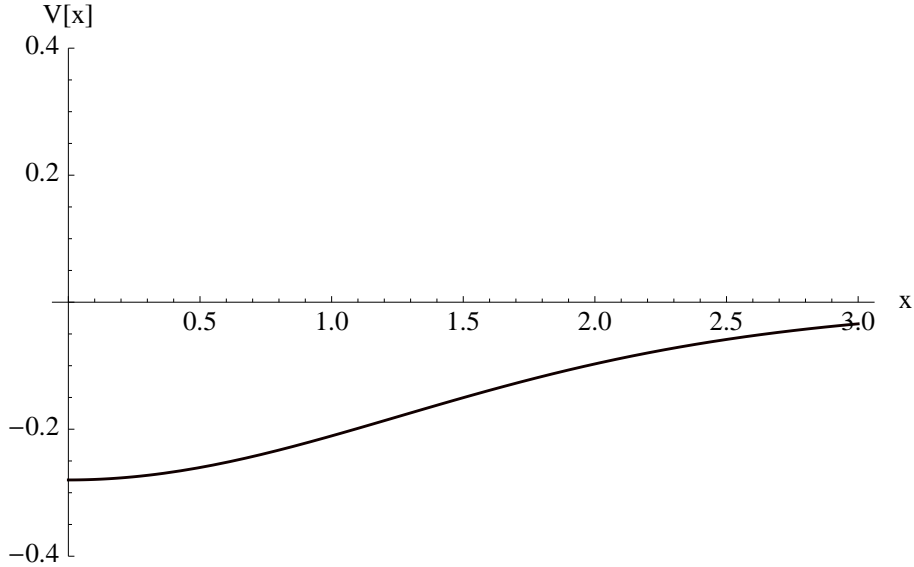}
\end{center}
\caption{\small Shape of the potential, Eq. (\ref{NCthreed185}).\label{fig2}}
\end{figure}

\section{Three-dimensional axionic electrodynamics}

We turn now to the problem of obtaining the interaction energy
between static point like sources for the three-dimensional version 
of the model studied in Ref. \cite{Gaete:2011ka}. To do this, we
shall start from the four-dimensional space-time Maxwell theory with
a term that couples the dual electromagnetic tensor to a fixed $v^\mu$
\cite{Carroll:1989vb,Goldhaber,Adam:2001ma,Kostelecky:2000mm,Casana:2008nw}:
\begin{equation}
{\cal L}^{\left( {3 + 1} \right)}  =  - \frac{1}{4}F_{\hat \mu \hat \nu } F^{\hat \mu \hat \nu } 
 + \frac{1}{4}\varepsilon ^{\hat \mu \hat \nu \hat \kappa \hat \lambda } v_{\hat \mu } A_{\hat \nu } F_{\hat \kappa \hat \lambda } 
 + \frac{1}{2}m^2 A_{\hat \mu } A^{\hat \mu }, \label{NCthreed190} 
\end{equation}
with the additional presence of a mass term for the gauge field. Here the
greek letters run from to $0$ to $3$. This model was considered in \cite{BaetaScarpelli:2003yd}, where the Proca mass stems from a Higgs scalar 
sector. It was shown that this model is unitary just for space like background 
while it presents ghost states for a timeline or lightlike background.

Next, to study this model in three-dimensional space-time dimensions, we
perform its dimensional reduction along the lines of \cite{Belich:2002vd}. In other words, we use the prescription: 
$A^{\hat \mu }  \to \left( {A^\mu  ;\phi } \right)$,
$v^{\hat \mu }  \to \left( {v^\mu  ;s} \right)$ and $\partial _3 \left( {anything}\right) = 0$. Carrying out this prescription in 
Eq. (\ref{NCthreed190}), we then obtain
\begin{eqnarray}
{\cal L}^{\left( {2 + 1} \right)}  &=&  - \frac{1}{4}F_{\mu \nu } F^{\mu \nu }  + \frac{1}{2}\partial _\mu  \phi \partial ^\mu  \phi  
+ \varepsilon ^{\mu \nu \lambda } v_\mu  \phi \left( {\partial _\nu  A_\lambda  } \right) + \frac{s}{2}\varepsilon ^{\mu \nu \lambda } A_\nu  
\left( {\partial _\mu  A_\lambda  } \right) \nonumber\\
&+& \frac{{m^2 }}{2}A_\mu  A^\mu   - \frac{{m^2 }}{2}\phi ^2,  \label{NCthreed191}
\end{eqnarray}
where $\mu ,\nu ,\lambda  = 0,1,2$. Accordingly, there appear two scalars, that is, the scalar field $\phi$ that exhibits dynamics, 
and $s$, a constant scalar. Then, by discarding the scalar field $s$ and the mass term for the gauge field, we arrive at 
\begin{equation}
{\cal L} =  - \frac{1}{4}F_{\mu \nu }^2  + \frac{1}{2}\phi
\varepsilon ^{\mu \nu \lambda } v_\mu  F_{\nu \lambda }  +
\frac{1}{2}\partial _\mu  \phi \partial ^\mu  \phi  - \frac{{m^2
}}{2}\phi ^2, \label{NCthreed195}
\end{equation}
which represents the three-dimensional analog of the model studied
previously \cite{Gaete:2011ka}. In addition, a preliminary study of this model was considered in \cite{Gaete:2005ba}.

Following our earlier procedure \cite{Gaete:2011ka}, we restrict ourselves to static scalar fields, 
a consequence of this is that one may replace $\Delta \phi = -\nabla ^2\phi$, with $\Delta \equiv\partial _\mu
\partial ^\mu$. It also implies that, after performing the integration
over $\phi$, the induced effective Lagrangian density is given by
\begin{equation}
{\cal L} =  - \frac{1}{4}F_{\mu \nu }^2  - \frac{{1
}}{8}\varepsilon ^{\mu \nu \lambda } v_\mu  F_{\nu \lambda }
\frac{1}{{\nabla ^2  - m^2 }}\varepsilon ^{\sigma \gamma \beta }
v_\sigma  F_{\gamma \beta }. \label{NCthreed200}
\end{equation}
By introducing $V^{\nu \lambda}  \equiv \varepsilon ^{\mu \nu
\lambda } v_\mu $, expression (\ref{NCthreed200}) then becomes
\begin{equation}
{\cal L} =  - \frac{1}{4}F_{\mu \nu }^2  - \frac{{1}}{8}V^{\nu
\lambda } F_{\nu \lambda } \frac{1}{{\nabla ^2  - m^2 }}V^{\gamma
\beta } F_{\gamma \beta }. \label{NCthreed205}
\end{equation}
Notice that (\ref{NCthreed205}) has the same form as the
corresponding effective Lagrangian density in four-dimensional
spacetime. This gives us the starting point for the
examination of the effects of the $Lorentz$ violating background
on the interaction energy.

It is once again straightforward to apply the gauge-invariant formalism discussed in the preceding section in the
$V^{0i} \ne 0$ and $V^{ij}=0$ ($v_0=0$) case (referred to as the spacelike
background in what follows). In such a case the Lagrangian reads 
\begin{equation}
{\cal L} =  - \frac{1}{4}F_{\mu \nu }^2  - \frac{{1}}{2}V^{0i}
F_{0i} \frac{1}{{\nabla ^2  - m^2 }}V^{0k} F_{0k},
\label{NCthreed210}
\end{equation}
where $(\mu ,\nu  = 0,1,2)$ and $(i,k= 1,2)$.
This leads us to the canonical Hamiltonian 
\begin{equation}
H_C  = \int {d^2 } x\left\{ { - A_0 \left( {\partial _i \Pi ^i} \right) + \frac{1}{2}{\bf \Pi} ^2  + \frac{{1
}}{{2}}\frac{{\left( {{\bf V} \cdot {\bf \Pi} } \right)^2
}}{{\left( { \nabla ^2  - M^2 } \right)}} + \frac{1}{2} B^2 }
\right\}, \label{NCthreed220}
\end{equation}
where $M^2\equiv m^2  + V^2$ and $B$ is the magnetic field.
We skip all the technical details and refer to \cite{Gaete:2011ka} for them. The corresponding static potential for two opposite 
charges located at ${\bf y}$ and ${\bf y^\prime}$ turns out to be 
\begin{equation}
V =  - \frac{{q^2 }}{{2\pi }}e^{M^2 \theta } \left\{ {K_0 \left( {ML} \right) - \frac{1}{2}\int\limits_{{\raise0.5ex\hbox{$\scriptstyle L$}
\kern-0.1em/\kern-0.15em
\lower0.25ex\hbox{$\scriptstyle {2M\theta }$}}}^\infty  {dt} \frac{1}{t}e^{ - {\textstyle{{ML} \over 2}}\left( {t + {\textstyle{1 \over t}}}
 \right)} } \right\} + \frac{{q^2 m^2 e^{M^2 \theta } }}{{4M}}L, \label{NCthreed255}
\end{equation}
where $L\equiv|{\bf y}-{\bf {y^\prime}}|$.
Again, this result explicitly displays the effect of including a
smeared source in the form of an ultraviolet finite static potential.
It is interesting to note that the rotational symmetry is restored
in the resulting form of the potential, although the external
background breaks the isotropy of the problem in a manifest way.
It should be remarked that this feature is also shared by the
corresponding four-dimensional spacetime interaction energy.

Here, an interesting matter comes out. The result (\ref{NCthreed255}) 
agrees with that of Polyakov \cite{Polyakov} based on the monopole plasma mechanism, except that this result shows a regularized Bessel function.
In this way the above analysis reveals that, although both models are different, the physical content is identical in the short distance regime.
This behavior is also obtained in the context of the condensation of topological defects
\cite{Gaete:2005ht,Gaete:2006ss}.

\section{Concluding Comments}

To conclude, this work is a sequel to \cite{Gaete:2011ka}, where we have considered a three-dimensional extension of the recently proposed 
finite axionic electrodynamics. To do this, we have exploited a crucial point for understanding the physical content of gauge theories, namely, 
the correct identification of field degrees of freedom with observable quantities. Again, our  calculations involve no $\theta$ expansion at all 
and, as in \cite{Gaete:2011ka}, the above analysis displays the key role played by the new quantum of length.

It is worth mentioning that our result of Section $2.1$ is finite for a special combination of both ultraviolet and infrared regulators.
This is different to what happens in $(3+1)$-dimensions. Another interesting point we mention here is that, according to the results of the 
papers \cite{Marchetti,Marchetti2}, a spin-charge $SU(2)$ x $U(1)$ gauged (planar) model can be written down which
is exactly equivalent to the original t-J model \cite{Spalek} for strongly correlated electronic systems. In this treatment, 
the dynamics of both  the $SU(2)$ and $U(1)$ gauge potentials is described by Chern-Simons actions. 
Considering this frame, we believe that our approach to extract potentials between static sources could be pushed forward and it would be 
relevant to understand how to obtain a low-energy effective action for the self-generated gauge field of the residual $U(1)$ gauge interaction 
induced by the spin-charge separation \cite{Marchetti3}. Our study of Section $2.2$ could be extended to take into account this gauge-invariant 
scenario which describes interesting physical properties of the t-J model.

We further note that, based on the investigation we have pursued in our work, we are faced with the prospect of going deeper into the study of 
axionic-like electromagnetic models in $(2+1)$-D. We have here considered an axionic action that stems from the dimensional reduction of a 
Lorentz-symmetry violating electrodynamical action. The scalar $\phi$-fieldÊ that mixes with the electromagnetic field-strength is the 
$(2+1)$-D descent of the $A_{3}$ -component of the four-dimensional potential. We intend, as a step forward, to write down the 
$(2+1)$-dimensionally reduced version of the true 
$\phi F_{\mu \nu } {\widetilde F}^{\mu \nu }$ of $(3+1)$-D and, then, analyze the planar counterpart of the Witten effect \cite{Witten} 
and place our approach for calculating potentials in a situation such that we may study a sort of planar topological insulator 
\cite{Fu,Kane,Moore} as an axionic medium. 

Finally, it seems a challenging work to extend the above analysis to the non-Abelian case as well as to three-dimensional gravity. 
We expect to report on progress along these lines soon.

\ack
P. Gaete was supported in part by Fondecyt (Chile) grant 1080260. One of us (PG) wants to thank the Abdus Salam ICTP for hospitality and 
support.  (J.H.-N.) expresses his gratitude to CNPq-Brasil and FAPERJ-Rio de
Janeiro for the invaluable support.\\
We would also like to thank Dr. W. Bietenholz for his useful comments and suggestions.\\

\end{document}